\documentclass[english,american,pra]{revtex4}
\usepackage{amsmath}
\usepackage{graphicx}

\makeatletter

\pdfpageheight\paperheight
\pdfpagewidth\paperwidth

\@ifundefined{textcolor}{}
{%
 \definecolor{BLACK}{gray}{0}
 \definecolor{WHITE}{gray}{1}
 \definecolor{RED}{rgb}{1,0,0}
 \definecolor{GREEN}{rgb}{0,1,0}
 \definecolor{BLUE}{rgb}{0,0,1}
 \definecolor{CYAN}{cmyk}{1,0,0,0}
 \definecolor{MAGENTA}{cmyk}{0,1,0,0}
 \definecolor{YELLOW}{cmyk}{0,0,1,0}
}

\usepackage{amsfonts}
\usepackage[american]{babel}
\usepackage{subfigure}
\usepackage{multirow}
\usepackage{color}

\newcommand{\bra}[1]{{\left\langle{#1}\right\vert}}
\newcommand{\ket}[1]{{\left\vert{#1}\right\rangle}}

\newcommand{\bracket}[2]{\langle#1|#2\rangle}
\newcommand{\comment}[1]{}

\makeatother

\usepackage{babel}
\begin{document}

\title{One-Dimensional Coinless Quantum Walks}

\author{Renato Portugal}
\affiliation{Laborat\'orio Nacional de Computa\c{c}\~ao Cient{\'\i}fica, Petr\'opolis, RJ
25651-075; Brazil}
\author{Stefan Boettcher and Stefan Falkner}
\affiliation{Department of Physics, Emory University, Atlanta, GA 30322; USA}

\begin{abstract}
A coinless, discrete-time quantum walk possesses a Hilbert space whose
dimension is smaller compared to the widely-studied coined
walk. Coined walks require the direct product of the site 
basis with the coin space, coinless walks operate purely in the site
basis, which is clearly minimal. These coinless quantum walks have
received considerable attention recently because they have evolution operators
that can be obtained by a graphical
method based on lattice tessellations
and they have been shown to be as efficient as the best known coined walks 
when used as a quantum search algorithm.  We argue that both formulations in their most general form are equivalent. In particular, we demonstrate how to transform the one-dimensional version of the coinless quantum 
walk into an equivalent extended coined version for a specific family of evolution operators.
 We present some of its basic, asymptotic features for 
the one-dimensional lattice with some examples of tessellations,
and analyze the mixing time and limiting
probability distributions on cycles.
\end{abstract}
\maketitle

\section{Introduction\label{sec:Introduction}}

The coined quantum walk (QW) on a line was introduced by 
Aharonov \textit{et al.}~\cite{Aharonov93} and
its generalization on regular graphs was studied in Ref.~\cite{AAKV01}. 
In this model, the particle hops from site to site 
depending on the value of an internal state of the walker, 
which plays the role of the coin. QWs on the line and on
multi-dimensional lattices spread quadratically
faster, in terms of the probability distribution, compared to the classical random 
walk model on the same underlying structure~\cite{AF0X}. 
The coined model was successfully applied to develop quantum searching algorithms, 
especially for finding a marked node in graphs~\cite{SKW03,AKR05,PortugalBook}.
These searching algorithms generalize Grover's algorithm in the sense
that the data is distributed spatially and a price must be paid 
to go from one place to another when searching some item.
The coined model in a regular graph of degree $d$ employs 
a $d$-dimensional unitary matrix that artificially
enlarges the Hilbert space and thus, memory space, in any quantum
computation. For non-regular graphs, the coined model requires
different coins depending on the degree of the site.

The success of the coined model has stimulated the study of alternative  QW models 
and a coinless version was introduced by Patel \textit{et al.}~\cite{Patel05}, and a coinless QW as a special case of a quantized cellular automaton even goes back to Meyer~\cite{Meyer96}. In Ref.~\cite{Patel05}, the authors construct the propagator in terms of two non-commuting unitary matrices, motivated by
 the staggered fermion formalism of quantum field theories.
It converts Dirac spinors used in continuous spacetime into spatial degrees
of freedom on discrete lattices, amounting to a new version of QWs 
with no internal degree of freedom, \textit{i.e.}, no coin. The authors
also applied their model to search a marked vertex in 
a two-dimensional lattice using numerical implementations~\cite{Patel12}.
Interestingly, the coinless model was rediscovered by 
Falk~\citep{Fal13}, who suggested a simple
method of obtaining the two-stroke propagator without 
the internal degrees of freedom by spliting the vertices of
the graph into disjoint patches that tessellate the two-dimensional
lattice. His method can be readily used for other graphs, such
as honeycombs and trees. 
Falk has also applied his model to search a marked vertex in 
the two-dimensional lattice using numerical implementations.
A rigorous analytical proof based on an asymptotic
analysis in terms of the system size, showing that the time complexity
of Falk's algorithm matches other QW models,
appeared in Ref.~\cite{APN13}. The advantage of this model
relies on using a smaller Hilbert space compared with
other discrete time models.
Recently, Ref.~\cite{CG14} used these ideas to improve the
efficiency of searching algorithms based on continuous-time 
QW models~\cite{PhysRevA.58.915} for 2D-lattices. The coinless model  
may also provide an alternative that could  be easier -- or at least cheaper -- to implement experimentally.

In this paper, we explore the
one-dimensional versions of the coinless model for both, the infinite line and a finite-sized cycle. We
present the coined and coinless models in a formalism that makes
explicit the connection between these models.  On the
line we use the saddle-point method to obtain the asymptotic (large $t$) 
probability distribution.
On the cycle, we obtain an explicit expression for the limiting 
probability distribution and present bounds for the mixing time ($\tau_{\epsilon}$).
Our numerical analysis suggests that $\tau_{\epsilon}=\Theta(N/\epsilon)$, where $N$
is the number of vertices.

This paper is organized as follows. 
In Sec.~\ref{sec:Typical-QW}, we introduce the 
coinless and coined QW models on one-dimensional lattices. 
In Sec.~\ref{sub:Equivalence-between-Coinless},
we present an example of the equivalence of both versions. 
In Sec.~\ref{sec:Abstract-QW}, we present an abstract formulation
that includes both QW models.
In Sec.~\ref{sec:Sample-Problems}, we present results for the one-dimensional coinless QW,  such as its   solution by Fourier analysis, the asymptotic form of its probability
density function, the Anderson-like localization for a coinless 3-state QW, and for the cycle we obtain an analytical expression for the
limiting probability distribution  and bounds for the mixing time.
In Sec.~\ref{sec:Conclusions}, we conclude with a discussion of the implications of our results.

\section{Typical formulations of Quantum Walks\label{sec:Typical-QW}}

The generic master-equation for any discrete-time quantum walk (DTQW), 
\begin{equation}
\left|\Psi\left(t+1\right)\right\rangle ={\cal U}\left|\Psi\left(t\right)\right\rangle ,\label{eq:MasterE}
\end{equation}
with the time-evolution operator (or propagator) ${\cal U}$ is formally
solved by $\left|\Psi\left(t\right)\right\rangle ={\cal U}^{t}\left|\Psi\left(0\right)\right\rangle $.
The Hilbert space of DTQW usually is a composite space which is spanned by the position states either augmented by  
coin states~\cite{AAKV01} or augmented by an auxiliary space the dimension of which is the same of the original position space~\cite{Szegedy04}. At the end of the evolution, the extra space is traced out to generate information about the position of the walker. Locality is demanded in the sense that the walker does only short jumps around their original positions.

The smallest conceivable Hilbert space for such a system is spanned
only by the site-basis $\left|\vec{n}\right\rangle $, also called the
computational basis, similar to one used in the continuous-time QW model~\cite{PhysRevA.58.915}. 
In such case, the state of the system can be described 
in terms of the site amplitudes $\psi_{\vec{x}}(t)=\left\langle \vec{x}|\Psi\left(t\right)\right\rangle $,
\textit{i.e.}, 
\begin{equation}
\psi_{\vec{x}}(t+1)=\sum_{\vec{y}}\left\langle \vec{x}\right|{\cal U}\left|\vec{y}\right\rangle \psi_{\vec{y}}(t).\label{eq:psint}
\end{equation}
 In a classical walk, the probability density function 
is simply given by $p_{\vec{x}}(t)=\psi_{\vec{x}}(t)$, but the fundamental
difference of any QW derives from the fact that $\psi_{\vec{x}}(t)$
represents a complex site-amplitude from which the physical density
is obtained via its modulus,
\begin{equation}
p_{\vec{x}}(t)=\left|\psi_{\vec{x}}(t)\right|^{2}=\left|\left\langle \vec{x}\right|{\cal U}^{t}\left|\Psi\left(0\right)\right\rangle \right|^{2}.\label{eq:PDF}
\end{equation}
Eq.~\eqref{eq:PDF}, by conservation of probability, implies that the evolution be unitary, and
thus, reversible and volume-preserving, while the classical process
is stochastic and irreversible and, hence, contractive. The modulus
of a superposition of (complex) amplitudes, in contrast with the mere addition of positive weights classically,
leads to a number of quantum effects, such as interference.

\begin{figure}[h!]
\begin{minipage}[b]{0.4\columnwidth}%
\includegraphics[angle=0,scale=0.35]{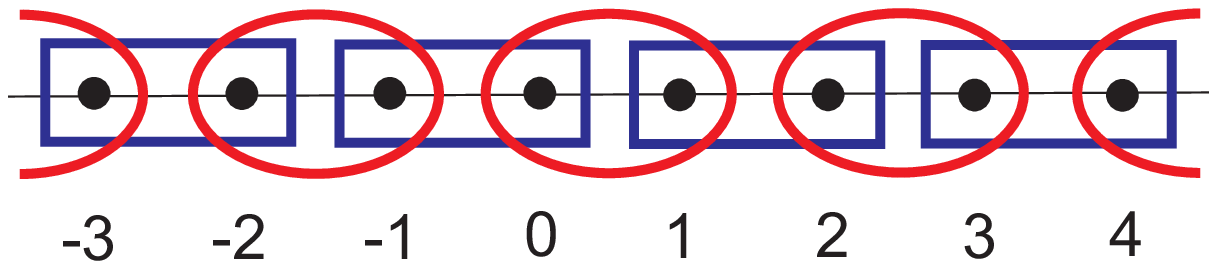}\vfill{\vspace{0.4cm}}
\includegraphics[scale=0.35]{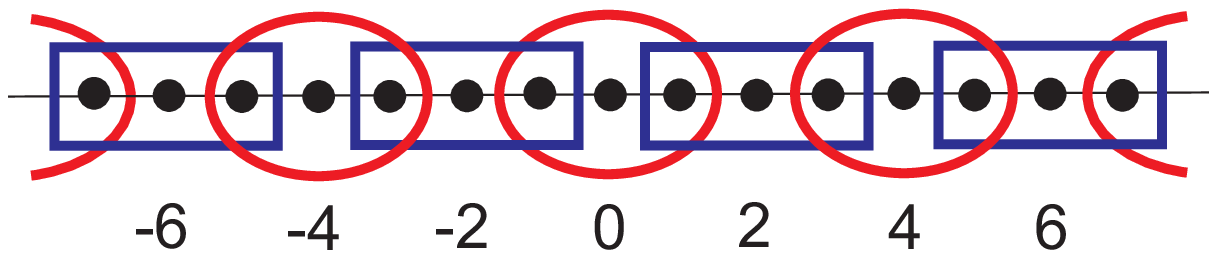}\vfill{\vspace{0.4cm}}
\end{minipage}\hfill{}%
\begin{minipage}[b]{0.48\columnwidth}%
\includegraphics[scale=0.30]{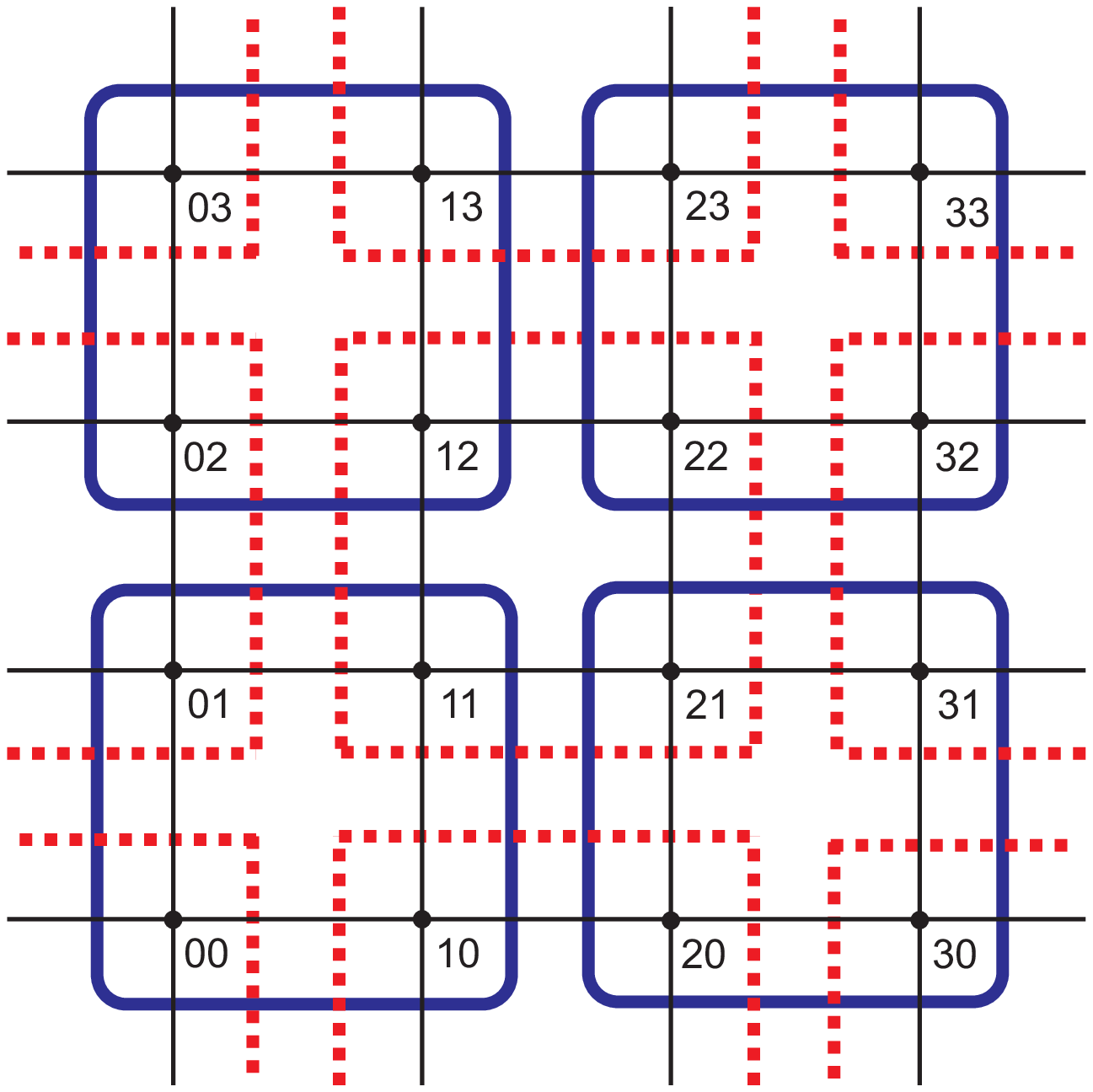}%
\end{minipage}\caption{\label{fig:Tesselations}Depiction of tessellation options for reflection
operators $U_{\vec{0},\vec{1}}$ in Eq.~(\ref{eq:FalkU}) in $D=1$
(left) and $D=2$ (right). Here, we use square blocks of length 2
in both dimensions (top left, and right). However, alternative (local) tessellations
are conceivable, such as block-length 3, as shown for $D=1$ (bottom
left),
which lead to equally efficient QW. As a minimal requirement, at least
the combination of both tessellations, needs
to cover the lattice and both should be sufficiently overlapping.}
\end{figure}

\subsection{Coinless Quantum Walk\label{sub:Coinless-QW}}

Patel \textit{et.al.}~\cite{Patel05} proposed a discrete-time quantum walk
with no auxiliary space, which is called coinless or staggered QW. This model was
re-invented in an alternative formulation in Ref.~\cite{Fal13}, which described an interesting method
to generate this kind of QW in a 2D-lattice by partitioning it into
disjoint patches that can be used to describe a reflection-based unitary
operator called $U_{\vec{0}}$ (or "even") operator. 
The full evolution operator is obtained after building a second partition,
which is analogous to the first one but diagonally shifted, generating a second
reflection-based unitary operator called $U_{\vec{1}}$ (or "odd") operator that does not commute with $U_{\vec{0}}$.
The resulting evolution operator then  is ${\cal U}=U_{\vec{1}}U_{\vec{0}}$
\footnote{Originally, Ref.~\citep{Fal13} introduced a quantum search algorithm
by interlacing a reflection  $U_{w}$ around the marked site $\left|w\right\rangle$ with the 
operators  $U_{\vec{0}}$ and $U_{\vec{1}}$, i.e.,   ${\cal U}=U_wU_{\vec{1}}U_wU_{\vec{0}}$.
If the reflection around the marked site is removed, the result is 
a propagator that generates the QW dynamics we study here.}.
One can obtain classes of evolution operators by choosing creative
graph tessellations that can considerably impact the effort needed to analyse the evolution. 

Falk's description can be generalized to a $D$-dimensional lattice
where the two-stroke reflection propagator consists of
\begin{equation}\label{eq:FalkU}
U_{\vec{0},\vec{1}}  =  2\sum_{\vec{x}}\left|u_{\vec{x}}^{\vec{0},\vec{1}}\right\rangle \left\langle u_{\vec{x}}^{\vec{0},\vec{1}}\right|-{\cal I},
\end{equation}
where the sum extends over all lattice sites $\vec{x}$ that have,
for $U_{\vec{0}}$, only even coordinates ($\vec{x}\equiv\vec{0}\mod2$)
or, for $U_{\vec{1}}$, only odd ones ($\vec{x}\equiv\vec{1}\mod2$).
Here, $\vec{0},\vec{1}$ refer to the binary vectors $\vec{0}=(0,\ldots,0)$
and $\vec{1}=(1,\ldots,1)$. The reflections $U_{\vec{0},\vec{1}}$
each mix the quantum state on hyper-square-blocks of $2^{D}$ sites
that cover the lattice, as shown in Fig.~\ref{fig:Tesselations}.
If we define $\vec{b}\in{\cal B}_{D}$, $0\leq b<2^{D}$, as that
vector on the $D$-dimensional hypercube ${\cal B}_{D}$ whose components
are the binary decomposition of the number $b$, we can write in general
\begin{equation}
\left|u_{\vec{x}}^{\vec{0},\vec{1}}\right\rangle =\sum_{\vec{b}\in{\cal B}_{D}}u_{\vec{b}}^{\vec{0},\vec{1}}\left|\vec{x}+\vec{b}\right\rangle ,\quad\left(\vec{x}\equiv\vec{0},\vec{1}\mod 2\right),\label{eq:FalkULA}
\end{equation}
with $\sum_{\vec{b}\in{\cal B}_{D}}\left|u_{\vec{b}}^{\vec{0},\vec{1}}\right|^{2}=1$,
which is a generalization of the operators introduced by Falk for
$D=2$ and restricted to uniform $u_{\vec{b}}^{\vec{0},\vec{1}}=\frac{1}{2}$.
The interlacing of $U_{\vec{0}}$ and $U_{\vec{1}}$ in the time evolution
${\cal U}^{t}\left|\Psi\left(0\right)\right\rangle $ spreads the
QW between diagonally offset, overlapping sets of blocks. Falk has
numerically analyzed the efficiency of a search algorithm based
on the two-dimensional version of this model. 
Ref.~\cite{APN13} rigorously proved that the efficiency
of Falk's algorithm matches other models of QWs.

As the most tractable case, we consider here the coinless QW
on the line, $D=1$, with the generalized form in Eq.~(\ref{eq:FalkULA}),
allowing for a tunable family of parameters $\alpha,\beta,\phi_1,\phi_2$
for the block-states: 
\begin{eqnarray}
\left|u_{x}^{{0}}\right\rangle  & = & \cos\frac{\alpha}{2}\left|2x\right\rangle +\textrm{e}^{i\phi_{1}}\sin\frac{\alpha}{2}\left|2x+1\right\rangle ,\label{eq:GenUAL_0}\\
\left|u_{x}^{{1}}\right\rangle  & = & \sin\frac{\beta}{2}\left|2x+1\right\rangle +\textrm{e}^{i\phi_{2}}\cos\frac{\beta}{2}\left|2x+2\right\rangle.\label{eq:GenUAL_1}
\end{eqnarray}
Here, $\left|u_{x}^{{0,1}}\right\rangle$ each cover blocks of merely two sites and the propagator is ${\cal U}=	U_\textrm{1}	U_\textrm{0}$, where
\begin{equation}\label{eq:1DcoinlessU}
U_\textrm{0,1} = 2\, \sum_{x=-\infty}^\infty \ket{u^\textrm{0,1}_{x}}\bra{u^\textrm{0,1}_{x}} - {\cal I}.
\end{equation}

It is now straightforward to calculate the matrix $\left\langle x\left|{\cal U}\right|y\right\rangle $
for the propagator ${\cal U}$ in the site basis by inserting Eqs.~(\ref{eq:GenUAL_0}) and~(\ref{eq:GenUAL_1}). Using 
\begin{eqnarray}
\left\langle x\biggm\vert u_{y}^{{0}}\right\rangle  & = & \cos\frac{\alpha}{2}\,\delta_{x,2y}+\textrm{e}^{i\phi_{1}}\sin\frac{\alpha}{2}\,\delta_{x,2y+1},\label{eq:Helpful}\\
\left\langle x\biggm\vert u_{y}^{{1}}\right\rangle  & = & \sin\frac{\beta}{2}\,\delta_{x,2y+1}+\textrm{e}^{i\phi_{2}}\cos\frac{\beta}{2}\,\delta_{x,2y+2},\nonumber 
\end{eqnarray}
and
\begin{equation}
\left\langle u_{x}^{{1}}\biggm\vert u_{y}^{{0}}\right\rangle  = \textrm{e}^{i\phi_{1}}\sin\frac{\alpha}{2}\sin\frac{\beta}{2}\,\delta_{x,y}+\textrm{e}^{-i\phi_{2}}\cos\frac{\alpha}{2}\cos\frac{\beta}{2}\,\delta_{x+1,y},\nonumber 
\end{equation}
to calculate $\left\langle 2x\left|{\cal U}\right|y\right\rangle$ and $\left\langle 2x+1\left|{\cal U}\right|y\right\rangle$, we obtain for the evolution equations
\begin{eqnarray}
\psi_{2x}(t+1) & = & \sin\alpha\sin\beta\,\textrm{e}^{i(\phi_{1}+\phi_{2})}\psi_{2x-2}(t)-\cos\alpha\sin\beta\,\textrm{e}^{i\phi_{2}}\psi_{2x-1}(t)-\nonumber \\
 &  & \cos\alpha\cos\beta\,\psi_{2x}(t)-\sin\alpha\cos\beta\,\textrm{e}^{-i\phi_{1}}\psi_{2x+1}(t),\label{eq:coinless2x}\\
\mbox{}\nonumber \\
\psi_{2x+1}(t+1) & = & \sin\alpha\cos\beta\,\textrm{e}^{i\phi_{1}}\psi_{2x}(t)-\cos\alpha\cos\beta\,\psi_{2x+1}(t)+\nonumber \\
 &  & \cos\alpha\sin\beta\,\textrm{e}^{-i\phi_{2}}\psi_{2x+2}(t)+\sin\alpha\sin\beta\,\textrm{e}^{-i(\phi_{1}+\phi_{2})}\psi_{2x+3}(t).\label{eq:coinless2xp1}
\end{eqnarray}

\subsection{Coined Quantum Walk\label{sub:Coined-QW}}

Due to its importance for quantum search algorithms, the most studied
formulation of a discrete-time QW proceeds by introducing a quantum
coin. In a coined QW, the time-evolution operator takes the form
\begin{equation}
{\cal U}={\cal S}\left({\cal C}\otimes{\cal I}\right),\label{eq:U=00003DSxC}
\end{equation}
containing the ``shift'' operator ${\cal S}$ and the quantum coin
${\cal C}$. Like the Bernoulli coin used to drive a classical random
walk, ${\cal C}$ is meant to determine which share of every on-site
amplitude gets transported to each of the neighboring sites. To preserve
unitarity, ${\cal C}$ is usually conceived of as a unitary operator
of a rank commensurate with the neighborhood degree $d$ of the site,
whose application entangles a $d$-dimensional \emph{vector} of site
amplitudes $\psi_{\vec{x}}(t)$ before the shift-operator ${\cal S}$
distributes those amplitudes at each site to its respective neighboring
sites. 

Such a coined QW possesses great conceptual clarity, and has been
shown to lead to the best-known efficiency for low-dimensional ($D\leq3$)
quantum search algorithms. However, the requirement to match the coin-space
to the neighborhood degree is not only a severe limitation to regular
lattices, but especially burdens the formulation with a Hilbert space
that is now the product of coin and site-space.

As a concrete example, let us discuss the nearest-neighbor QW on a
line, which only requires a unitary coin matrix of rank $d=2$, 
in its most general form given by 
\begin{equation}
{\cal C}=\left(\begin{array}{cc}
\cos\rho & \sin\rho\, {\textrm{e}}^{i\theta}\\
\sin\rho\, {\textrm{e}}^{i\varphi} & -\cos\rho\, {\textrm{e}}^{i\left(\theta+\varphi\right)}
\end{array}\right)\label{eq:QW1dCoin}
\end{equation}
in terms of three real parameters ($\rho,\theta,\varphi$).
It is conventional to expand the solutions in the coin- and the site
basis $\left|s\right\rangle \otimes\left|x\right\rangle =\left|s,x\right\rangle $,
where $s=0,1$ refers to the two directions out of each site, and
$x$ labels the sites on the line. The most general form of ${\cal U}$
on the 1D-line is then
\begin{eqnarray}
{\cal U} & = & \sum_{x}\sum_{\mu,\nu=0}^{1}\left\{ A_{\mu,\nu}\left|\mu,x+1\right\rangle \left\langle \nu,x\right|+B_{\mu,\nu}\left|\mu,x-1\right\rangle \left\langle \nu,x\right|\right\} ,\label{eq:propagator}
\end{eqnarray}
with matrices $A=P{\cal C}$ and $B=Q{\cal C}$, where
\begin{equation}
P=\left(\begin{array}{cc}
1 & 0\\
0 & 0
\end{array}\right),\quad Q=\left(\begin{array}{cc}
0 & 0\\
0 & 1
\end{array}\right),\label{eq:PQsimple}
\end{equation}
in a commonly used (but restrictive) interpretation of the shift operator at each site.
Taking, say, $\theta=\frac{\pi}{2}$ and $\varphi=0$ in Eq.~(\ref{eq:QW1dCoin}),
we get for the evolution equations in (\ref{eq:psint}) for the upper
and lower component $\psi_{x}^{(0,1)}(t)$ of the wave function:
\begin{eqnarray}
\psi_{x}^{(0)}(t+1) & = & \cos\rho\,\psi_{x+1}^{(0)}(t)-\sin\rho\,\psi_{x+1}^{(1)}(t),\nonumber \\
\psi_{x}^{(1)}(t+1) & = & \sin\rho\,\psi_{x-1}^{(0)}(t)+\cos\rho\,\psi_{x-1}^{(1)}(t).\label{eq:1d2state_psi}
\end{eqnarray}
These, or similar systems, have been studied at great length~\citep{Ambainis01,Bach2004562,PortugalBook}.
Occasionally, a coined QW with the possibility to remain at the same
site is considered, leading to a three-term propagator in Eq.~(\ref{eq:propagator})
\citep{IK05,Falkner14a}. However, in this formulation, such a walk
already requires a coin of rank $d=3$.

\section{Relation between Coinless and Coined Quantum Walk on a Line\label{sub:Equivalence-between-Coinless}}

The coinless evolution equations (\ref{eq:coinless2x}) and (\ref{eq:coinless2xp1}) can be rewritten as two-dimensional
vectors:
\begin{eqnarray}
\left[\begin{array}{c}
\psi_{2x}(t+1)\\
\psi_{2x+1}(t+1)
\end{array}\right] & = & A\left[\begin{array}{c}
\psi_{2x-2}(t)\\
\psi_{2x-1}(t)
\end{array}\right]+M\left[\begin{array}{c}
\psi_{2x}(t)\\
\psi_{2x+1}(t)
\end{array}\right]+B\left[\begin{array}{c}
\psi_{2x+2}(t)\\
\psi_{2x+3}(t)
\end{array}\right],\label{eq:coinlessPsi}
\end{eqnarray}
where we set
\begin{equation}
A=\sin\beta\,{\textrm{e}^{i(\phi_1+\phi_2)}} P{\cal C},\quad B=\sin\beta\,{\textrm{e}^{i(\phi_1+\phi_2)}} Q{\cal C}\quad M=\cos\beta\,{\textrm{e}^{i(\phi_1+\phi_2)}} R{\cal C}\label{eq:coinlessABM}
\end{equation}
with $R=\left[\begin{array}{cc}
0 & -{\textrm{e}^{i\phi_2}}\\
{\textrm{e}^{-i\phi_2}} & 0
\end{array}\right]$ and $P,Q$ as in Eq.~(\ref{eq:PQsimple}), and the coin ${\cal C}$
in Eq.~(\ref{eq:QW1dCoin}) for $\rho=\frac{\pi}{2}-\alpha$, $\theta=\pi-\phi_1$ 
and $\varphi=-(\phi_1+2\phi_2)$.

We note that these equations are very similar to those for the coined
QW in Eq.~(\ref{eq:1d2state_psi}), if we reinterpret consecutive
pairs of an even site and the odd site to its right into a single
node by treating the two original sites as upper and lower coin states, respectively.
This mapping is exact only for $\beta=\frac{\pi}{2}$, for which the
self-term with $M$ in Eq.~(\ref{eq:coinlessPsi}) disappears. For other values of $\beta$, we require a generalized interpretation of the shift operators. This
corresponds to the symmetric choice for $\left|u_{x}^{{1}}\right\rangle $
in Eq.~(\ref{eq:GenUAL_1}), however, the equally symmetric choice of
$\alpha=\frac{\pi}{2}$ for $\left|u_{x}^{{0}}\right\rangle $
provides only a degenerate, one-sided QW with ${\cal C}\to{\cal I}$.
The equivalent of the Hadamard QW would only emerge for the choice
$\alpha=\frac{\pi}{4}$, since Eq.~(\ref{eq:coinlessPsi}) is formally
equivalent to Eq.~(\ref{eq:1d2state_psi}) when $\rho=\frac{\pi}{4}$. 
Still, it is remarkable that the meaning
of the two reflection operators $U_{0}$ and $U_{1}$
of the coinless QW in Eq.~(\ref{eq:1DcoinlessU}) in this case can be 
related to corresponding sequence of applying coin ${\cal C}$
and shift operator ${\cal S}$ in the coined QW. In general, however,
the possibilities of the definition for the coinless QW in 
Eqs.~(\ref{eq:GenUAL_0}) and~(\ref{eq:GenUAL_1})
exceed those provided for by the conventional coined QW as presented
in Sec.~\ref{sub:Coined-QW}. We therefore provide an abstract
formulation that unifies both, coined and coinless QW, in a more general
framework.

\section{Generalized Quantum Walks\label{sec:Abstract-QW}}

As the preceding discussion shows, the coined and coinless QW possess
very similar structures. The sequence of coin and shift operation closely resembles the action of the two-stroke reflection operator. Yet, each already presents a very specific choice  of operations (shifts, reflections, etc) that are suggested by their intuitive nature for their respective situations. Due to those restrictions, their equivalence can be shown only under certain circumstances, as we have discussed in Sec.~\ref{sub:Equivalence-between-Coinless}. Removing those restrictions reveals a more general equivalence, which we explore in the following. 

The master equation, Eq.~(\ref{eq:MasterE}), merely imposes unitartity, however, to obtain a physical walk we also demand locality over a bounded neighborhood (i.e., a sparse adjacency matrix), which we here simply take as nearest-neighbor exchanges. We then define on a $d$-regular network, such as a lattice with $d=2D$, the propagator 
\begin{eqnarray}
{\cal U} & = & \sum_{\vec{n}}^{{\cal N}/r}\left\{ M\left|\vec{n}\right\rangle \left\langle \vec{n}\right|+\sum_{\mu=1}^{d}A_{\mu}\left|\vec{n}\right\rangle \left\langle \vec{n}+\vec{e}_{\mu}\right|\right\} ,\label{eq:genU}
\end{eqnarray}
over a partial basis $\left\{ \left|\vec{n}\right\rangle \right\}$
of size ${\cal N}/r$ in the Hilbert space of rank ${\cal N}$.
$A_{\mu}$ or $M$ are operators of rank $r$ that determine the portion
of the wave function at site $\vec{n}$ that either gets transported
to one of $d$ neighboring site $\vec{n}+\vec{e}_{\mu}$ or remains
at the site during the next update, respectively. Here, it could be
${\cal N}=rN$ for the product of rank-$r$ coin and site space in the coined
QW with $\vec{n}$ denoting all sites. Alternatively, ${\cal N}=N$
and the ${\cal N}/r$ subset of sites $\vec{n}$ labels blocks possessing
$r$ block-internal sites, such as in Eq. (\ref{eq:FalkULA}), each
in some tessellation of the network in the coinless QW. In that
case, $M$ entangles those $r$ sites in each block anckered at $\vec{n}$,
while the $A_{\mu}$ transfers amplitudes to the $d$ adjacent blocks
labeled by $\vec{n}+\vec{e}_{\mu}$. For our one-dimensional example
in Sec. \ref{sec:Typical-QW}, the former leads to evolution equations
such as in (\ref{eq:1d2state_psi}), while the latter leads to Eq. (\ref{eq:coinlessPsi}).
Note that in Eq. (\ref{eq:genU}) we required a certain level of \textit{(i)}
homogeneity (in that each block has the same operators $M,A_{\mu}$
and hence the same number of adjacent blocks, $d$), and \textit{(ii)} reciprocity
(in that there exist pairs $\mu,\nu=\nu\left(\mu\right)$ such that
for every two adjacent blocks $\vec{n},\vec{m}$ with $\vec{n}+\vec{e}_{\mu}=\vec{m}$
it is also $\vec{m}+\vec{e}_{\nu}=\vec{n}$). Both conditions are obvious
on a uniform $D$-dimensional rectangular lattice with $r=d=2D$ neighboring
blocks, where simply $\vec{e}_{\nu}=-\vec{e}_{\mu}$ for each direction,
but requires some thought for more general networks. 

Using  Eq.~(\ref{eq:genU}) and expanding ${\cal U}^{\dagger}{\cal U}$, the unitarity conditions implies that
\begin{eqnarray}
{\cal I}_{r} & = & M^{\dagger}M+\sum_{\mu=1}^{d}A_{\mu}^{\dagger}A_{\mu},\nonumber \\
0 & = & A_{\mu}^{\dagger}M+M^{\dagger}A_{\nu\left(\mu\right)},\quad\left(1\leq\mu\leq d\right)\nonumber \\
0 & = & A_{\mu}^{\dagger}A_{\nu},\quad\left(1\leq\mu,\nu\leq d,\mu\not=\nu\right),\label{eq:unitarityCond}
\end{eqnarray}
which implies that $M+\sum_{\mu=1}^{d}A_{\mu}$ must be unitary.
These conditions can not be satisfies in general by scalars, 
except for trivial cases. Thus, similar in spirit to Dirac's derivation
of relativistic QM, this algebra requires
matrix respresentations of the hopping operators and multi-component spinors to represent states. 

One specific representation is provided by $r$-dimensional hopping matrices with $r\geq d$ that arise from the combination of coin- and shift matrices, such as for the line in Eq.~(\ref{eq:propagator}), with ${\cal N}=Nr$. Another specific representation results from the combination of operators, as those in Eq. ~(\ref{eq:1DcoinlessU}), in the coinless case that lead to relations between neighboring sites on a patch, such as Eqs.~(\ref{eq:coinlessPsi}-\ref{eq:coinlessABM}) for the coinless QW on the line. The latter has the apparent advantage that site-space and Hilbert space coincide, ${\cal N}=N$, however, every site within a patch of size $r$ requires a separate considertion. As we will see in Sec.~\ref{sec:Sample-Problems},  for example, the coinless walk on a line requires a  Fourier analysis that is staggered between even and odd sites for $r=d=2$.

More general choices are possible than the traditional coined walk,
with specific shift operators for the transfer of a single component
of the wave-function along a lattice direction, or the reflection
operators for our coinless QW. Also, the algebra in Eq. (\ref{eq:unitarityCond})
can also be satisfied by a higher-dimensional vector space than the degree imposed by the network, $r\geq d$. We have
not tried here to enumerate systematically the most general choice even for the simple
line, as most would likely not lead to any new phenomena. Note, however,
that already adding a third component to the QW ($r=3$) on the line ($d=2D=2$), coined~\citep{IK05,Falkner14a}
or coinless (see Sec. \ref{sub:A-coinless-3-state} below), allows
for localization effects that do not exist for any two-dimensional
operators.

\section{Results for Coinless Quantum Walks on a Line\label{sec:Sample-Problems}}

In this Section, we list a few equivalent results for properties
of the one-dimensional coinless QW that mirror those that have been
found important for the understanding of a coined QW. First, we provide
explicit results for the wave-function on even and odd sites of the
infinite line as Fourier integrals. From these, we determine the asymptotic
probability density function (PDF). We also discuss alternative tessellations,
involving three sites simultaneously, that are equivalent to the behavior
for a three-state coined QW, but seem to exhibit a
more varied localization property.  
In the next Section, we consider the coinless
QW on a finite-length cycle, and use its solution to provide bounds
on the mixing time.

\subsection{Fourier Solution on the Infinite Line\label{sub:Fourier-Solution-on}}

We define as a Fourier transform, staggered between even and odd sites,
the vectors
\begin{eqnarray}
\ket{\tilde{\psi}_{k}^{\textrm{0}}} & = & \sum_{x=-\infty}^{\infty}\textrm{e}^{-2xki}\ket{2x},\label{eq:Pie}\\
\ket{\tilde{\psi}_{k}^{\textrm{1}}} & = & \sum_{x=-\infty}^{\infty}\textrm{e}^{-(2x+1)ki}\ket{2x+1},\label{eq:Pio}
\end{eqnarray}
where $k\in[-\pi,\pi]$. For a fixed $k$, they define a plane that
is invariant under the action of the evolution operator. The analysis
of the dynamics can be reduced to a two-dimensional subspace of ${\cal H}$
by defining a reduced evolution operator 
\begin{equation}
U_{\textrm{RED}}^{(k)}=\left[\begin{array}{cc}
A & -B^{*}\\
B & A^{*}
\end{array}\right],
\end{equation}
where 
\begin{eqnarray}
A & = & -\cos\alpha\cos\beta+\sin\alpha\sin\beta\,\textrm{e}^{i(\phi_{1}+\phi_{2})}\textrm{e}^{2ki},\label{eq:A}\\
B & = & \sin\alpha\cos\beta\,\textrm{e}^{i\phi_{1}}\textrm{e}^{ki}+\cos\alpha\sin\beta\,\textrm{e}^{-i\phi_{2}}\textrm{e}^{-ki}.\label{eq:B}
\end{eqnarray}
$U_{\textrm{RED}}^{(k)}$ is unitary because $A\, A^{*}+B\, B^{*}=1.$
The eigenvalues are $\lambda=\textrm{e}^{\pm i\theta}$ and can be
obtained from the characteristic polynomial, which is ${\lambda}^{2}-(A+A^{*})\lambda+1.$
Then 
\begin{equation}\label{eq:costheta}
\cos\theta=\frac{A+A^{*}}{2}.
\end{equation}
The quantities defined in these expressions depend on $k$ and 
others parameters. We will make explicit those parameters only in some equations.

There are trivial solutions that are obtained by taking either $A=0$
or $B=0$. $U_{\textrm{RED}}^{(k)}$ in the latter case is diagonal
and the wave function moves to right or to the left without spreading.
On the other hand, if $A=0$, the wave function oscillates back and
forth without spreading. The first case is obtained when $(\alpha,\beta)=(0,0)$,
$(\pi/2,\pi/2)$ or by adding any multiple of $\pi$ to $\alpha$
or $\beta$; and the second case when $(\alpha,\beta)=(0,\pi/2)$
or by adding any multiple of $\pi$. In any of those cases, the eigenvalues
of the evolution operator is evenly spread out on the unit complex
circle.

The non-trivial eigenvectors of $U_{\textrm{RED}}^{(k)}$ are 
\begin{equation}
\frac{1}{\sqrt{C^{\pm}}}\left(\begin{array}{c}
-B^{*}\\
\textrm{e}^{\pm i\theta}-A
\end{array}\right),\label{eq:eigenvec}
\end{equation}
where 
\begin{equation}
C^{\pm}=\sin\theta\,\big(2\sin\theta\pm i\,(A-A^{*})\big).
\end{equation}
There are special values of $k$ such that $C^{\pm}=0$. They are:
1) $\alpha=\beta$ and $\phi_{1}+\phi_{2}+2k=\pm\pi$, 2) $\alpha+\beta=\pi$
and $\phi_{1}+\phi_{2}+2k=0,\pm2\pi$. In those cases, we have to
take the limit of eigenvectors (\ref{eq:eigenvec}) when $k$ approach
to the troublesome values.

The eigenvectors of the full propagator $U$ associated with eigenvalues $\textrm{e}^{\pm i\theta}$
are 
\begin{equation}
\ket{v_{k}^{\pm}}=\frac{1}{\sqrt{C^{\pm}}}\left(-B^{*}\ket{\tilde{\psi}_{k}^{\textrm{0}}}+(\textrm{e}^{\pm i\theta}-A)\ket{\tilde{\psi}_{k}^{\textrm{1}}}\right),
\label{eq:Uev}
\end{equation}
and we can write
\begin{equation}
U=\int_{-\pi}^{\pi}\frac{\textrm{d}k}{2\pi}\left(\textrm{e}^{i\theta}\ket{v_{k}^{+}}\bra{v_{k}^{+}}+\textrm{e}^{-i\theta}\ket{v_{k}^{-}}\bra{v_{k}^{-}}\right).\label{eq:U}
\end{equation}

If we take $\ket{\psi(0)}=\ket{0}$ as initial condition, the walker's state at
time $t$ is 
\begin{eqnarray}
\ket{\psi(t)} & = & \sum_{x=-\infty}^{\infty}\left(\psi_{2x}(t)\,\ket{2x}+\psi_{2x+1}(t)\,\ket{2x+1}\right),
\end{eqnarray}
where 
\begin{equation}
\psi_{2x}(t)=\int_{-\pi}^{\pi}\frac{\textrm{d}k}{2\pi}\,|B|^{2}\left(\frac{\textrm{e}^{i(\theta t-2kx)}}{C^{+}}+\frac{\textrm{e}^{-i(\theta t+2kx)}}{C^{-}}\right)\label{eq:psike}
\end{equation}
and 
\begin{equation}
\psi_{2x+1}(t)=\int_{-\pi}^{\pi}\frac{\textrm{d}k}{2\pi}\,\frac{B\sin\theta t}{\sin\theta}\,\textrm{e}^{-(2x+1)ki}.\label{eq:psiko}
\end{equation}
The probability distribution is obtained after calculating 
$p_{2x}(t)=\left|\psi_{2x}(t)\right|^2$ and
$p_{2x+1}(t)=\left|\psi_{2x+1}(t)\right|^2$. In the next Section, we
use the saddle-point expansion method to obtain an asymptotic expression for 
the  probability distribution.  The probability distribution
is not symmetric in this case. An alternative initial condition which 
yields a symmetric QW is $(\ket{0}+i\ket{1})/2$.

\subsection{Asymptotic PDF of a coinless Quantum Walk\label{sub:Asymptotics-of-a}}

The probability density function (PDF) $p_{{x}}(t)$ for a walker
to be at a site ${x}$ at time $t$ provides the most comprehensive
description of any QW. All of its properties, such as the mean-square
displacement or first-return probabilities, derive from the PDF, making
it the central object of any textbook discussion of classical random
walks~\citep{Itzykson89,Redner01} and transport processes~\citep{Weiss94}.
This also holds for QWs~\citep{PortugalBook,VA12}. 
%
%Considering the similarities between coined and coinless
%QW demonstrated in Sec.~\ref{sub:Equivalence-between-Coinless}, it is no surprise
%that its PDF for the coinless case
%exhibits the well-known features familiar from the coined QW~\citep{Ambainis01},
%see for example also Ref.~\citep{Patel05}. However, the PDF for a
%coinless QW differs in subtle ways, for instance, on how the initial
%conditions enter the calculation, see Sec.~\ref{sub:Fourier-Solution-on}. 

Starting from Eq.~\eqref{eq:psike} {[}or, equivalently, Eq.~\eqref{eq:psiko}{]},
we define an effective Hamiltonian, 
\begin{equation}
{\cal H}_{\pm}\left(k\right)=-2vk\pm\theta(k),\label{eq:Hamiltonian}
\end{equation}
and pursue a stationary-phase solution for $t\gg1$~\citep{BO} via
$\frac{\partial}{\partial k}{\cal H}_{\pm}\left(k_{\pm}\right)=0$,
\textit{i.e.}, 
\begin{equation}
v=\frac{x}{2t}=\pm\frac{1}{2}\frac{\partial\theta\big(k_{\pm}\big)}{\partial k}.\label{eq:stationarypoint}
\end{equation}
Note that it is convenient to introduce the maximal speed $2$, such
that the effective velocity $|v|<1$ is gauged in these units, obtained
from the furthest reach in a given tessellation of a single application
of $U$, which here is uniformly the side-length of a 2-block. Finally,
expanding $ $${\cal H}_{\pm}\left(k\right)\sim{\cal H}_{\pm}\left(k_{\pm}\right)+\frac{1}{2}{\cal H}_{\pm}^{\prime\prime}\left(k_{\pm}\right)\left(k-k_{\pm}\right)^{2}$
and substituting $k=k_{\pm}+(\pm1+i)u$, with $u$ as the new integration
variable, allows the asymptotic evaluation of the integral in Eqs.~(\ref{eq:psike}) 
and~(\ref{eq:psiko}) as a Gaussian along a complex contour. 

\begin{figure}
\includegraphics[trim=1cm 15cm 1cm 1cm,scale=0.5]{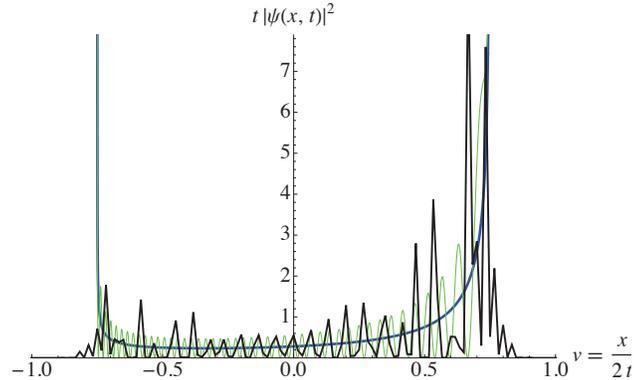}%%%bb=100bp 415bp 560bp 730bp,clip,
\caption{\label{fig:1dPDFnocoin}Plot of the rescaled PDF $tp_{x}(t)\sim\rho\left(v\right)$
at $t=30$ in the scaling variable $v=\frac{x}{2t}$ for the coinless
QW in $D=1$ at $\alpha+\beta=\pi$ and $v_{0}=\sin\alpha=\frac{3}{4}$,
initiated at $t=0$  on site $x=0$, $\left|\Psi\left(0\right)\right\rangle =\left|0\right\rangle $.
The simulation (dark line) contains some more structure than the full
asymptotic form (green shaded line) from Eq.~(\ref{eq:psi_10}) can
represent at $t=30$. The envelope function (thickened blue line),
also obtained from Eq.~(\ref{eq:psi_10}) by removing the explicitly $t$-dependent phase, represents the average behavior
and eccentricity of the exact data quite well, within the range of
its validity bracketed by poles at $v=\pm v_{0}$.}
\end{figure}

For example, for $\alpha+\beta=\pi$ and defining the ``effective''
velocity $v_{0}=\sin\alpha$, \textit{i.e.}, 
$\theta_{\pm}(k)=\mp\arccos\left(1-2v_{0}^{2}\sin^{2}k\right)$
from Eq.~(\ref{eq:costheta}), we find the solutions of Eq.~(\ref{eq:stationarypoint})
\begin{equation}
k_{\pm}=\arccos\left[\pm\frac{v}{v_{0}}\sqrt{\frac{1-v_{0}^{2}}{1-v^{2}}}\right],\label{eq:k0}
\end{equation}
which are valid only for $\left|v\right|<v_{0}$. Hence, only for
$\alpha=\beta=\frac{\pi}{2}$, \textit{i.e.}, uniform {$u_{x}^{\vec{0},\vec{1}}=\frac{1}{\sqrt{2}}$}
in Eq.~(\ref{eq:FalkULA}), can the 1D-QW reach maximal speed $c$,
however, at that point the walk degenerates into one-sided lockstep
motion. At $k_{\pm}$, we find 
\begin{equation}\label{eq:SPvalues}
{\cal H}_{\pm} = 2vk_{\pm}\mp\arccos\left[\frac{1+v^{2}-2v_{0}^{2}}{1-v^{2}}\right],
\end{equation}
and
\begin{equation}
\left|{\cal H}_{\pm}^{\prime\prime}\left(k_{\pm}\right)\right| = \left(1-v^{2}\right)\sqrt{\frac{v_{0}^{2}-v^{2}}{1-v_{0}^{2}}}.
\end{equation}
Replacing the values for $k_{\pm}$ into the expressions $|B|^{2}/C^{\pm}$ inside the integrals (\ref{eq:psike}) and~(\ref{eq:psiko}), working out all cases we eventually
%\begin{equation}
%\frac{|B|^{2}}{C^{\pm}} =
%\frac{1}{2}\left[\begin{array}{cc}
%              1+v & \mp\sqrt{1-v^{2}} \\
%\mp\sqrt{1-v^{2}} & 1-v
%\end{array}\right].
%\end{equation}
find for Eq.~(\ref{eq:PDF}) for odd sites: 
\begin{equation}
p_{2x+1}(t)  \sim \frac{1}{\pi}\,\sqrt{\frac{1-v_{0}^{2}}{v_{0}^{2}-v^{2}}}\,\cos^{2}\left\{ \frac{\pi}{4}+t\,{\cal H}_{\pm}\big(k_{\pm}\big)\right\} \label{eq:psi_10}
\end{equation}
and $p_{2x}(t)\sim\frac{1+v}{\left(1-v\right)}p_{2x+1}(t)$ for
$\left|v\right|<v_{0}<1$
%, where $\beta\sim\frac{(2x)}{2t}=\frac{x}{t}$
at $t\gg1$. We plot the resulting PDF as $p_{\vec{x}}(t)=\left(p_{2x}(t)+p_{2x+1}(t)\right)/2$
in Fig.~\ref{fig:1dPDFnocoin} and compare it with exact simulations.
Note the asymmetry of the PDF; for $v_{0}\to1$ ($\alpha\to \pi/2$)
it degenerates into $\rho(v)\to\delta\left(v_{0}\mp v\right)$
for $\left|\Psi\left(0\right)\right\rangle =\left|0\right\rangle $.

In the $2D$ case, the real eigenvalues of the reduced propagator produce
a motionless delta spike at the origin generating Anderson localizations,
similar to the ones that have been noticed on the coined QW model~\citep{IK05,Falkner14a}.
The $1D$ case analyzed in this paper has no static localization and
the reduced propagator does not have real (unit) eigenvalues, except
for trivial limiting cases. An easy way to introduce localization
in $1D$ coined QWs is by enlarging the coin space, see for example the
models addressed in~\citep{IK05,Falkner14a} and our discussion in
Sec.~\ref{sub:A-coinless-3-state}. In the tessellation model, it
seems that the blocks need to have at least three sites in order to
produce localization.

\subsection{A coinless 3-state QW\label{sub:A-coinless-3-state}}

Here, we reconsider a one-dimensional coinless QW for the 
tessellation shown on the bottom-left of Fig.~\ref{fig:Tesselations}. 
This form has a few interesting new features not previously considered for a
coinless QW:~\textit{(i)} It can be shown that every pair of reflection operators generically
leads to localization.~\textit{(ii)} In this tessellation, neither set of
blocks, for the first nor the second reflection operator, completely
covers the lattice individually, however, the combination of both
sets of blocks do, and they have an overlap that is symmetric and
reaches every part of the lattice.%

For consistency in the notation with Eq.~\eqref{eq:FalkU}, we define
the evolution operator as ${\cal U} = U_{1}U_{0}$ with
the two reflection operators as
\begin{equation}
U_{0,1} = 2\sum_{x=0}^\infty\left|u_{x}^{0,1}\right\rangle \left\langle u_{x}^{0,1}\right|-{\cal I},
\end{equation}
with 
\begin{eqnarray}
\left|u_{x}^{\textrm{0}}\right\rangle  & = & \frac{1}{\sqrt 3}\left(\left|4x-1\right\rangle +\left|4x\right\rangle +\left|4x+1\right\rangle\right),\\
\left|u_{x}^{\textrm{1}}\right\rangle  & = & \frac{1}{\sqrt 3}\left(\left|4x+1\right\rangle +\left|4x+2\right\rangle +\left|4x+3\right\rangle\right).
\end{eqnarray}
The staggered Fourier basis is spanned by four vectors
\begin{equation}
	\ket{\tilde{\psi}_{k}^{j}} = \sum_{x=-\infty}^{\infty}\textrm{e}^{-(4x+j)ki}\ket{2x+j},\label{eq:psitilde3}
\end{equation}
where $0\le j \le 3$. The four-dimensional space spanned by those vectors is invariant
under the action of the evolution operator and allows to use a reduced evolution operator
given by
\begin{equation}
	U_{\textrm{RED}}=\frac{1}{9}\left[ \begin {array}{cccc} 3&-6\,{{\rm e}^{-ik}}&0&-6\,{{\rm e}^{ik}
}\\ \noalign{\medskip}4\,{{\rm e}^{-3 ik}}-2\,{{\rm e}^{ik}}&4\,{
{\rm e}^{-4 ik}}+1&-6\,{{\rm e}^{-ik}}&-2\,{{\rm e}^{-2 ik}}-2\,{
{\rm e}^{2 ik}}\\ \noalign{\medskip}4\,{{\rm e}^{-2 ik}}+4\,{{\rm e}
^{2 ik}}&4\,{{\rm e}^{-3 ik}}-2\,{{\rm e}^{ik}}&3&-2\,{{\rm e}^{-ik}
}+4\,{{\rm e}^{3 ik}}\\ \noalign{\medskip}-2\,{{\rm e}^{-ik}}+4\,{
{\rm e}^{3 ik}}&-2\,{{\rm e}^{-2 ik}}-2\,{{\rm e}^{2 ik}}&-6\,{
{\rm e}^{ik}}&1+4\,{{\rm e}^{4 ik}}\end {array} \right], 
\end{equation}
such that
\begin{equation}
	{\cal U}\,\ket{\tilde{\psi}_{k}^{j}}=\sum_{l=0}^3 \left\langle l\left|U_{\textrm{RED}}\right|j\right\rangle \ket{\tilde{\psi}_{k}^{j}}.
\end{equation}

\begin{figure}
\centering \includegraphics[scale=0.45]{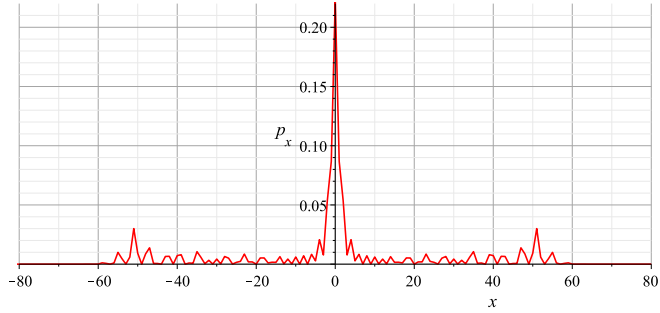} 
\caption{Probability distribution of the coinless 3-state QW after 20 steps
with initial state located at $x=0$. The sharp peak at the origin is 
the signature of the locality always present in the 3-state case.
\label{fig:coinless3state}}
\end{figure}

The eigenvalues of $U_{\textrm{RED}}$ are 1 with multiplicity 2 and $\textrm{e}^{\pm i \theta}$, where
\begin{equation}
	\cos \theta = \frac{1}{9}\left( 4\cos 4k - 5 \right).
\end{equation}
The fact that $U_{\textrm{RED}}$ has eigenvalue 1 (which does not depend on $k$) implies that
this QW has localization, that is, part of the wave function does not move. 
Fig.~\ref{fig:coinless3state} shows the probability distribution of the walk after 20 steps
with initial state located at $x=0$. The sharp peak at the origin does not move and is
always present there regardless the number of steps.
This is in contrast to the coined 3-state walk, where there
is no localization for the Fourier coin, for example.

\section{Coinless QW on the Cycle\label{sec:cycle}}

Consider a cycle with even number $N$ of sites. The associated Hilbert
space ${\cal H}^{N}$ is spanned by $\{\ket{x},0\le x\le N-1\}$.
Some of the equations of Sec.~\ref{sub:Fourier-Solution-on} apply here after
a formal replacement $\exp(i)\rightarrow\omega_N$, where 
$\omega_{N}=\textrm{e}^{2\pi i/N}$.
Eqs.~(\ref{eq:Pie}) and (\ref{eq:Pio}) must be changed so that
the dummy index $x$ runs from 0 to $N/2-1$.

The Fourier transform is based on the following set of orthonormal
vectors: 
\begin{eqnarray}
\ket{\tilde{\psi}_{k}^{\textrm{0}}} & = & \sqrt{\frac{2}{N}}\sum_{x=0}^{\frac{N}{2}-1}\omega_{N}^{-2xk}\ket{2x},\\
\ket{\tilde{\psi}_{k}^{\textrm{1}}} & = & \sqrt{\frac{2}{N}}\sum_{x=0}^{\frac{N}{2}-1}\omega_{N}^{-(2x+1)k}\ket{2x+1},
\end{eqnarray}
where $0\le k\le N/2-1$.
Eqs.~(\ref{eq:A}) and (\ref{eq:B}) must be changed to 
\begin{eqnarray}
A & = & -\cos\alpha\cos\beta+\sin\alpha\sin\beta\,\textrm{e}^{i(\phi_{1}+\phi_{2})}\omega_N^{2k},\label{eq:A_N}\\
B & = & \sin\alpha\cos\beta\,\textrm{e}^{i\phi_{1}}\omega_N^{k}+\cos\alpha\sin\beta\,\textrm{e}^{-i\phi_{2}}\omega_N^{-k},\label{eq:B_N}
\end{eqnarray}
and Eq.~(\ref{eq:U}) to 
\begin{equation}
U=\sum_{k=0}^{\frac{N}{2}-1}\left(\textrm{e}^{i\theta}\ket{v_{k}^{+}}\bra{v_{k}^{+}}+\textrm{e}^{-i\theta}\ket{v_{k}^{-}}\bra{v_{k}^{-}}\right),\label{eq:U_N}
\end{equation}
Eqs.~(\ref{eq:psike}) and (\ref{eq:psiko}) to 
\begin{equation}
\psi_{2x}(t)=\sqrt{\frac{2}{N}}\sum_{x=0}^{\frac{N}{2}-1}|B|^{2}\omega_N^{-2xk}\left(\frac{\textrm{e}^{i\theta t}}{C^{+}}+\frac{\textrm{e}^{-i\theta t}}{C^{-}}\right)\label{eq:psike_N}
\end{equation}
and 
\begin{equation}
\psi_{2x+1}(t)=\sqrt{\frac{2}{N}}\sum_{x=0}^{\frac{N}{2}-1}\frac{B\sin\theta t}{\sin\theta}\,\omega_N^{-(2x+1)k}.\label{eq:psiko_N}
\end{equation}
If 4 divides $N$ and $\alpha=\beta$ and $\phi_{1}=\phi_{2}$, 
eigenvector~(\ref{eq:eigenvec})
and the normalization constants $C^{\pm}$ are zero for $k=N/4$.
The eigenvectors for this special value of $k$ must be calculated separately.

\subsection{Mixing Time\label{sub:Mixing-Time}}

Let $\ket{v_{k}}$ and $\textrm{e}^{i\lambda_{k}}$ denote the eigenvectors
and eigenvalues of ${\cal U}$. Then 
\begin{equation}
{\cal U}=\sum_{k=0}^{N-1}\textrm{e}^{i\lambda_{k}}\ket{v_{k}}\bra{v_{k}}.\label{eq:Ulambda}
\end{equation}
The \textit{time-averaged} probability density function (PDF) is defined by 
\begin{equation}
\overline{p}_{x}(T)=\frac{1}{T}\sum_{t=0}^{T-1}p_{x}(t),
\end{equation}
where by Eq.~(\ref{eq:Ulambda}),
we obtain for the PDF on  a finite-length cycle
\begin{equation}
\overline{p}_{x}(T)=\sum_{k,k'=0}^{N-1}c_{k}^{*}c_{k'}v_{k,x}v_{k',x}^{*}\,\frac{1}{T}\sum_{t=0}^{T-1}\textrm{e}^{i(\lambda_{k}-\lambda_{k'})T},
\end{equation}
setting $c_{k}=\bracket{0}{v_{k}}$ and $v_{k,x}=\bracket{x}{v_{k}}$
for a walker that departs from the origin. With
\begin{equation}
\frac{1}{T}\sum_{t=0}^{T-1}\textrm{e}^{i(\lambda_{k}-\lambda_{k'})T}=\delta_{\lambda_{k},\lambda_{k'}}+(1-\delta_{\lambda_{k},\lambda_{k'}})\frac{\textrm{e}^{i(\lambda_{k}-\lambda_{k'})T}-1}{T(\textrm{e}^{i(\lambda_{k}-\lambda_{k'})}-1)}
\end{equation}
we obtain 
\begin{equation}
\overline{p}_{x}(T)=\pi_{x}+\frac{1}{T}\sum_{\begin{subarray}{c}
k,k'=0\\
\lambda_{k}\neq\lambda_{k'}
\end{subarray}}^{N-1}c_{k}^{*}c_{k'}v_{k,x}v_{k',x}^{*}\,\frac{\textrm{e}^{i(\lambda_{k}-\lambda_{k'})T}-1}{\textrm{e}^{i(\lambda_{k}-\lambda_{k'})}-1},
\end{equation}
with 
\begin{eqnarray}
\pi_{x} & := & \lim_{T\rightarrow\infty}\overline{p}_{x}(T)\,=\sum_{\begin{subarray}{c}
k,k'=0\\
\lambda_{k}=\lambda_{k'}
\end{subarray}}^{N-1}c_{k}^{*}c_{k'}v_{k,x}v_{k',x}^{*}.\label{eq:pi_x}
\end{eqnarray}

The mixing time $\tau_{\epsilon}$ is defined as 
\begin{equation}
\tau_{\epsilon}=\min\{T\,|\,\forall t\geq T,\, D(\overline{p}(t),\pi)\le\epsilon\},
\end{equation}
where $D(\overline{p}(t),\pi)$ is the total variation distance (TVD) defined by
\begin{equation}
D(\overline{p}(t),\pi)=\frac{1}{2t}\sum_{x=0}^{N-1}\left|\sum_{\begin{subarray}{c}
k,k'=0\\
\lambda_{k}\neq\lambda_{k'}
\end{subarray}}^{N-1}c_{k}^{*}c_{k'}v_{k,x}v_{k',x}^{*}\,\frac{\textrm{e}^{i(\lambda_{k}-\lambda_{k'})t}-1}{\textrm{e}^{i(\lambda_{k}-\lambda_{k'})}-1}\right|.\label{eq:Dpbarpi}
\end{equation}
The mixing time captures the notion of the convergence of the time-averaged density to the limiting PDF.

\subsubsection{Time-Averaged Probability Density Function in the Large-Time Limit\label{sub:Limiting-Probability-Distributio}}

In analogy with the discussion leading to the eigenvalues of the propagator on the infinite line in Eq.~(\ref{eq:Uev}), the coefficients  $v_{k,x}$ are given by 
\begin{eqnarray}
v_{k,2x} & = & \frac{-\sqrt{2}B^{*}}{\sqrt{NC^{\pm}}}\,\omega_N^{-2xk}\\
v_{k,2x+1} & = & \frac{\sqrt{2}\,(\textrm{e}^{\pm i\theta}-A)}{\sqrt{NC^{\pm}}}\,\omega_N^{-(2x+1)k}
\end{eqnarray}
and $c_{k}=v_{k,0}$, where the sign plus must be taken for $0\le k<N/2$
and the sign minus for $N/2\le k<N$.
Using the fact that $\lambda_{k}=\lambda_{k'}$ when $k'=k$ and $k'=N/2-k$,
and Eq.~(\ref{eq:pi_x}), the limiting PDF for even sites
is 
\begin{equation}
\pi_{2x}=\frac{2}{N^{2}}+\frac{4}{N^{2}}\sum_{k=1}^{\frac{N}{2}-1}|B|^{4}\left(\frac{1}{{C^{+}}^{2}}+\frac{1}{{C^{-}}^{2}}+\frac{2\cos\frac{8\pi k\, x}{N}}{C^{+}C^{-}}\,(1-\delta_{k\frac{N}{4}})\right).
\end{equation}
Notice that the last term vanishes when $k=N/4$. For odd sites, the
limiting probability is 
\begin{equation}
\pi_{2x+1}=\frac{2}{N^{2}}\sum_{k=0}^{\frac{N}{2}-1}\frac{|B|^{2}}{\sin^{2}\theta}+\frac{1}{N^{2}}\sum_{k=1}^{\frac{N}{2}-1}\frac{1-\delta_{k\frac{N}{4}}}{\sin^{2}\theta}\left(B^{2}\omega_{N}^{-2k(2x+1)}+{B^{*}}^{2}\omega_{N}^{2k(2x+1)}\right).
\end{equation}

\begin{figure}[h!]
\centering \includegraphics[scale=0.35]{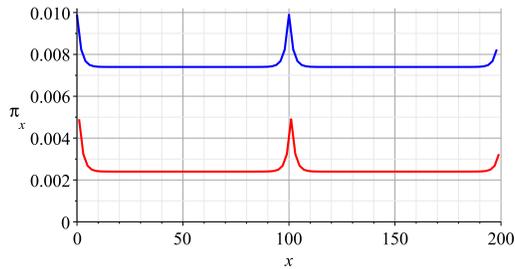} 
\caption{Limiting probability density function for $N=200$. Upper blue (lower red) curve is the PDF for even (odd) sites. 
\label{fig:limiting_prob}}
\end{figure}

Fig.~\ref{fig:limiting_prob} shows the limiting PDF for even
and odd sites separately (the parameters 
are $N=200$, $\alpha=\pi/2$, $\beta=2\pi/3$, and $\phi_{1}=\phi_{2}=0$). Both PDF are almost constant except for the spikes
around $x=0$ and $x=N/2$. If 4 does not divide $N$, there are no spikes at $x=N/2$.

If is interesting to compare those results with the corresponding ones
for the coined QW~\cite{AAKV01,PortugalBook,BGKLW03}. In both cases, the
limiting PDFs are remarkably similar. If $N$ is even, the 
limiting PDF for the coined case also have spikes at $x=0$ and $x=N/2$.
On the other hand, if $N$ is odd, the limiting PDF for the coined
case is constant for all $x$ while there is no simple tessellation in the
coinless case that allows the definition of an interesting evolution operator
in this case.

\subsubsection{Asymptotic Behavior of the Mixing Time\label{sub:Asymptotic-Behavior-of}}

From Eq.~(\ref{eq:Dpbarpi}), we see that that the TVD 
between the average and the limiting PDF depends
on $t$ in two distinct ways: 1) it decreases linearly with $t$ because
the pre-factor $1/t$, 2) it oscillates because the term $\textrm{e}^{i(\lambda_{k}-\lambda_{k'})t}$.
The oscillatory term makes difficult the exact determination of $\tau_{\epsilon}$
in terms of $\epsilon$ and the system size $N$. We use the Riemann-Lebesgue
lemma to argue that this oscillatory term can be disregarded to calculate
the asymptotic scaling behavior of the mixing time.

\begin{figure}[h!]
\centering \includegraphics[scale=0.45]{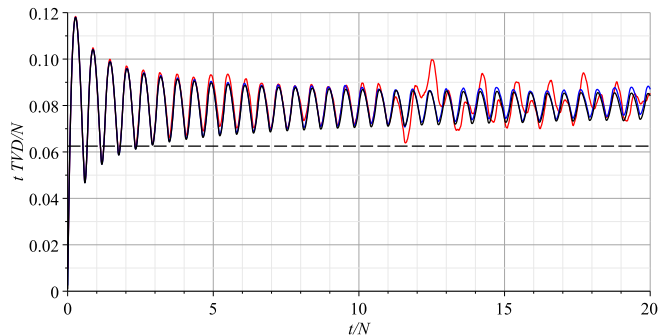} 
\caption{Rescaled total variation distance ($t\,\textrm{TVD}/N$) as function
of $t/N$ for $N=500,2000,5000$ with colors red, blue, and black, respectively.
The dashed line corresponds to the TVD without the oscillatory term,
\textit{i.e.} after removing the term $\exp(i(\lambda_{k}-\lambda_{k'})t)$.
\label{fig:TVD_rescaled}}
\end{figure}

Fig.~\ref{fig:TVD_rescaled} shows the rescaled graphs of $tD(\overline{p}(t),\pi)/N$
as a function of $t/N$ for three different values of $N$ (the parameters 
are $\alpha=\pi/2$, $\beta=2\pi/3$, and $\phi_{1}=\phi_{2}=0$).
The curves are strikingly coincident for small $t/N$ and gradationally
becomes more distinguishable as $t/N$ increases because the curves
have different amplitudes while still keeping similar wavelengths. The same
coincidence happens for larger values of $N$ that we can verify numerically.
 This numerical result implies that, for large $N$, 
$D(\overline{p}(t),\pi)$ has the form $\frac{N}{t}f\big(\frac{t}{N}\big)$,
where $f\big(\frac{t}{N}\big)$ is an oscillatory function the amplitude of which
does not depend on $N$.
We argue based on the Riemann-Lebesgue lemma that the amplitude of
the oscillation dies off when $t/N$ gets larger.

From Eq.~(\ref{eq:Dpbarpi}) we obtain the following bound for the TVD 
\begin{equation}
D(\overline{p}(t),\pi)\le\frac{1}{2t}\sum_{x=0}^{N-1}\left(\left|\sum_{k,k'}f_{k,k',x}\right|+\left|\sum_{k,k'}f_{k,k',x}\textrm{e}^{i\Delta_{k,k'}t}\right|\right),\label{eq:upperbound}
\end{equation}
where the sums over $k,k'$ are the same as before (i.e., such that $\lambda_{k} \not= \lambda_{k'}$), $\Delta_{k,k'}=\lambda_{k}-\lambda_{k'}$
and 
\begin{equation}
f_{k,k',x}\,=\,\frac{c_{k}^{*}c_{k'}v_{k,x}v_{k',x}^{*}}{\textrm{e}^{i(\lambda_{k}-\lambda_{k'})}-1}.
\end{equation}
For large $N$, the second term of Eq.~(\ref{eq:upperbound}) can
be approximated by a double integral and the Riemann-Lebesgue lemma
states that it goes to zero when $t\rightarrow\infty$. (There may be a small neighborhood in the integration domain where $\Delta_{k,k'}\sim 1/N$ such that the lemma would not apply, however, the weight of this contribution is negligible.)  This 
proves that $\tau_{\epsilon}$ is $\Theta(1/\epsilon)$ for any fixed
$N$. 

By analysing the sign of $f_{k,k',x}$, we can simplify the
first term of (\ref{eq:upperbound}), which is given by 
\begin{eqnarray}\label{eq:termf}
\sum_{x=0}^{N-1}\left|\sum_{k,k'}f_{k,k',x}\right| & = & \frac{4}{N^{2}}\sum_{\begin{subarray}{c}
k,k'=0\\
k\neq k'\\
k+k'\neq\frac{N}{2}
\end{subarray}}^{\frac{N}{2}-1}\frac{\left|B_{k}B_{k'}\right|^{2}}{C_{k}^{+}C_{k'}^{+}}\,(1-g^-_{k,k'})+\nonumber \\
 &  & \frac{4}{N^{2}}\sum_{k,k'}^{\frac{N}{2}-1}\frac{\left|B_{k}B_{k'}\right|^{2}}{C_{k}^{+}C_{k'}^{-}}\,(1+(\delta_{k,k'}-1)\, g^+_{k,k'})+\nonumber \\
 &  & \frac{2}{N}\sum_{k}^{\frac{N}{2}-1}\frac{\left|B_{k}\right|^{2}}{{C_{k}^{+}}{C_{k}^{-}}}{(\textrm{e}^{i\theta_{k}}-A_{k})(\textrm{e}^{-i\theta_{k}}-A_{k})},
\end{eqnarray}
where 
\begin{equation}
g^\pm_{k,k'}\,=\,\frac{(-1)^{k-k'}-\cos\frac{2\pi(k-k')}{N}}{\sin\frac{2\pi(k-k')}{N}}\,\frac{1+\cos(\theta_{k} \pm \theta_{k'})}{\sin(\theta_{k} \pm \theta_{k'})}.
\end{equation}

\begin{figure}[h!]
\centering \includegraphics[scale=0.35]{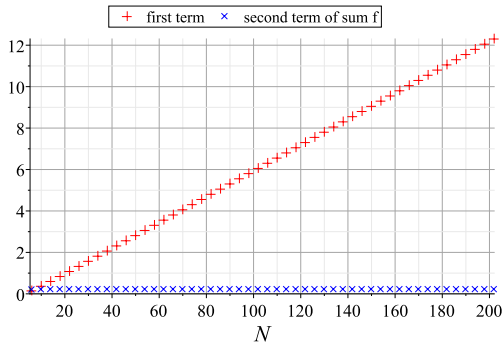} 
\caption{Behavior of the first (red $+$) and second (blue $\times$)
terms of  Eq.~(\ref{eq:termf}) as a function of $N$. The first term
clearly is $O(N)$.
\label{fig:mixingtime_termf}}
\end{figure}

Fig.~\ref{fig:mixingtime_termf} shows the plot of the first (red $+$) 
and second (blue $\times$) terms of the rhs of Eq.~(\ref{eq:termf}) as a function of $N$.
The third term has no poles and can be calculated numerically by converting
the sum into an integral, which is independent of $N$ for large $N$. This result,
together with Fig.~\ref{fig:TVD_rescaled}, numerically
shows that the TVD is $\Theta(N)$. Then $\tau_{\epsilon}=\Theta(N/\epsilon)$,
that is, $\tau_{\epsilon}$ is propotional to $N/\epsilon$ for $N$
large enough and $\epsilon$ small enough.
Notice that the corresponding result for the coined QW on 
cycles (with odd $N$) is the bound $O(N\log N/\epsilon^3)$ found in
Ref.~\cite{AAKV01}.

\section{Conclusions\label{sec:Conclusions}}

In conclusion, we have shown a formal connection between the
coinless and coined one-dimensional QW,
obtained a number of asymptotic properties for coinless QWs
on infinite 1D-lattice, showed Anderson-like localization in
coinless 3-state models, and analyzed the asymptotic behavior of the
mixing times on cycles. The numerical results suggest that 
$\tau_{\epsilon}=\Theta(N/\epsilon)$. 

A comparison between the coinless and
the coined QWs reveals striking similarities. For instance, the two-step
propagator combined with a initial conditions localized inside the
cell that contains the origin produce asymptotic behavior similar
to coined QWs departing from the origin with generic initial coin
state. And just as for the coined QW, where the $2x2$ Grover coin degenerates
in $D=1$ to make a non-trivial \emph{symmetric} QW at optimal speed
$c$ impossible, we find here that the $D=1$ tessellation in 
Eqs.~(\ref{eq:GenUAL_0}) and~(\ref{eq:GenUAL_1}) has only trivial solutions for $v_{0}\to1$ while
in $D=2$ and the 3-state model it provides both symmetry and optimal speed. Finally, we
note that using square-block tessellations of size $2^{D}$ in $D$
dimensions becomes impractical for $D>2$, as the rank of the reduced
space grows exponentially as $2^{D}$ while the degree of each site
only grows $\sim2D$. That suggests that better tessellations would
need to be found, such as a $D$-dimensional version of ``3-crosses'',
as a generalization of  the 3-patches in Fig.~\ref{fig:Tesselations}.

The nature of coinless QWs depend on the choice of the tessellation.
We have used simple 2-blocks neighboring sites which cover the entire
lattice. Many other tessellations are possible, as we have demonstrated
for the 3-site tessellation for one-dimensional QW as showed in
Fig.~\ref{fig:Tesselations}, for example. Note that in this case
neither of the two-stroke tessellation individually covers the entire
lattice, only the combination of both tessellations. We believe that
a symmetric and overlapping tessellation with a combined coverage
is sufficient for an optimal coinless quantum search. It will be interesting
to analyze in more detail which kind of physical behavior the choice
of a tessellation can provide, such as for displacement, localization
\citep{Inui05,Falkner14a}, and mixing. This physical analysis helps
to understand algorithmic applications of this QW model,
such as searching for a marked vertex.

\section*{Acknowledgements}

We acknowledge financial support from CNPq, Faperj, and the U. S. National Science Foundation through grant DMR-1207431. SB thanks LNCC for its hospitality and acknowledges financial support through a research fellowship through the “Ciência sem Fronteiras” program in Brazil.

%\bibliographystyle{apsrev}
%\bibliography{Boettcher}

\end{document}